\documentclass[]{pasj01}
\Accepted{$\langle$acception date$\rangle$}
\usepackage{graphicx}
\usepackage{color}
\usepackage{ifthen}
\usepackage{xcolor}
\usepackage{natbib}

\begin{document}

\title{Discovery of a strong 6.6 keV emission feature from EXO 1745$-$248 after the superburst in 2011 October}
\author{Wataru B. Iwakiri\altaffilmark{1,2,}$^{*}$}
\altaffiltext{1}{Department of Physics, Faculty of Science and Engineering, Chuo University, 1-13-27 Kasuga, Bunkyo-ku, Tokyo 112-8551, Japan}
\altaffiltext{2}{RIKEN, 2-1 Hirosawa, Wako, Saitama 351-0198, Japan}
\author{Motoko Serino\altaffilmark{3}}
\altaffiltext{3}{Department of Physical Sciences, Aoyama Gakuin University,
5-10-1 Fuchinobe, Chuo-ku, Sagamihara, Kanagawa 252-5258, Japan}
\author{Tatehiro Mihara\altaffilmark{2}}
\author{Liyi Gu\altaffilmark{2}}
\author{Hiroya Yamaguchi\altaffilmark{4}}
\altaffiltext{4}{Japan Aerospace Exploration Agency, Institute of Space and Astronautical Science,Chuo-ku, Sagamihara, Kanagawa 252-5210, Japan}
\author{Megumi Shidatsu\altaffilmark{5}}
\altaffiltext{5}{Department of Physics, Ehime University, 2-5, Bunkyocho, Matsuyama, Ehime 790-8577, Japan}
\author{Kazuo Makishima\altaffilmark{2,6}}
\altaffiltext{6}{Kavli Institute for the Physics and Mathematics of the Universe, The University of Tokyo, 5-1-5 Kashiwa-no-ha, Chiba 277-8583,Japan}
\email{iwakiri@phys.chuo-u.ac.jp}


\maketitle
%

\begin{abstract}
We discover an unidentified strong emission feature in the X-ray spectrum of EXO 1745$-$248 obtained by RXTE at 40 hr after the peak of a superburst. The structure was centered at 6.6 keV and significantly broadened with a large equivalent width of 4.3 keV, corresponding to a line photon flux of 4.7 $\times$ 10$^{-3}$ ph cm$^{-2}$ s$^{-1}$. The 3--20 keV spectrum was reproduced successfully by a power law continuum with narrow and broad (2.7 keV in FWHM) Gaussian emission components. Alternatively, the feature can be described by four narrow Gaussians, centered at 5.5 keV, 6.5 keV, 7.5 keV and 8.6 keV. Considering the strength and shape of the feature, it is unlikely to have originated from reflection of the continuum X-rays by some optically thick materials, such as an accretion disk. Moreover, the intensity of the emission structure decreased significantly with an exponential time scale of 1 hr.
The feature was not detected in an INTEGRAL observation performed 10 h before the RXTE observation with a line flux upper limit of 1.5 $\times$ 10$^{-3}$ ph cm$^{-2}$ s$^{-1}$. The observed emission structure is consistent with gravitationally redshifted charge exchange emission from Ti, Cr, Fe, and Co. We suggest that the emission results from a charge exchange interaction between a highly metal-enriched fall back ionized burst wind and an accretion disk, at a distance of $\sim$60 km from the neutron star. If this interpretation is correct, the results provide new information on the understanding of nuclear burning processes during thermonuclear X-ray bursts.

\end{abstract}


\section{Introduction}
An X-ray emitting binary called a Neutron-star Low-mass X-ray Binary (NS-LMXB) consists of a mass-donating low-mass star, and a mass-accreting neutron star (NS) with weak magnetic fields. These NSs often produce X-ray bursts by unstable thermonuclear burning of accreted H/He \citep[e.g.,][]{fujimoto81}, and are hence thought to be a factory of elements up to or heavier than iron \citep[e.g.,][]{wallace81,schatz01}. Since such heavy-element ashes sink beneath an optically thick atmosphere on the NS surface, they will not be easily detected in astronomical observations. Recently, however, several spectral features as signs of highly enriched heavy elements have been reported. For example, \citet{kubota19} found a spectral feature at 30 keV from Aquila X-1 (a typical NS-LMXB with recurrent outbursts) in the decay phase of an outburst, and argued that it is a K-shell structure of some heavy elements synthesized via the rapid proton capture process. Moreover, indications of highly enriched heavy elements were found as significant spectral absorption features in burst spectra \citep{intzand10,barriere2015,iwai2017}. Theoretical studies of \citet{weinberg06} and \citet{Yu2018} also predict that some heavy-element ashes are ejected by radiation-driven winds during the so-called photospheric radius expansion (PRE) phase of X-ray bursts.

The spectral features suggestive of burst-produced heavy elements were also detected in a superburst spectrum from 4U 1820$-$30 \citep{strohmayer02, intzand10}. Superbursts are X-ray bursts with unusually long ($\geq$1000 s) durations, and their origin is thought to be carbon flashes \citep{cumming2001,strohmayer02} in sources with accretion rates above 10\% of Eddington rate.  Because the estimated ignition column depths of superbursts,  $\sim$10$^{12}$ g cm$^{-2}$ \citep[e.g.,][]{cumming06}, are higher than those of normal type-I X-ray bursts ($\sim 10^8$ g cm$^{-2}$), a larger amount of heavy elements under the surface of a bursting NS could be ejected by a superburst than by a normal type-I X-ray burst. The problem is that even if heavy elements deposited on the surface of a NS are ejected or exposed by a superburst in a source with a high accretion rate, it is difficult to obtain their information from X-ray spectra because of the bright persistent emission from the accretion disk. In recent years, however, MAXI observations have revealed that superbursts can occur (though not frequently) even in objects with low accretion rates \citep{serino16}. One possible origin of
superbursts from sources with such low accretion rates, where the nuclear burning of H/He does not produce enough carbon for nuclear reactions, is thought to be the ignition of a thick He layer \citep{kuulkers2010,intzand17}. Since such superbursts may allow us to observe signs of heavy elements ejected by the strong burst wind due to He ignition without being buried under their persistent emission, we are encouraged to search X-ray spectra after the superbursts for evidence of heavy elements.

\begin{table*}
  \tbl{Log of the RXTE Observations of EXO 1745$-$248.}{%
  \begin{tabular}{lccccc}
  \hline\noalign{\vskip3pt} 
  \hline\noalign{\vskip3pt}             
  Label &  Obs ID & Obs Time &  & PCA \\ 
  &    (96316-01-)  & Start / End (MJD)  & Exp.(ks)         & Rate$^{\dagger}$ (cps)               \\ 
  \hline
  A  & 40-00       & 55860.207 / 55860.251    & 1.73           &14.16$ \pm$0.12     \\
  B   & 40-01       & 55860.971 / 55860.994    & 1.78           & 157.1 $ \pm$0.4     \\
  C   & 40-02       & 55861.040 / 55861.065    & 2.06           & 169.2$ \pm$0.4    \\
  D  & 41-00       & 55863.470 / 55863.492   & 1.17           & 64.51$ \pm$0.26    \\
  E  & 41-01       & 55864.517 / 55864.533    & 0.98           & 9.90$ \pm$0.14    \\
  F  & 41-02       & 55865.145 / 55865.212    & 3.18           & 10.85$ \pm$0.08    \\
   \hline
\end{tabular}}
\label{table1}
\begin{tabnote}
$^{\dagger}$ Background subtracted count rate with PCU2 only, in 3--20 keV.
\end{tabnote}
\end{table*}
On 2011 October 24, the Monitor of All-sky X-ray Image (MAXI) \citep{matsu09} detected a very long superburst from the direction of the globular cluster Terzan 5 \citep{altamirano12, serino12}. The source was identified with the X-ray source EXO 1745$-$248 by a Chandra observation on 2011 November 3 \citep{pooley11,altamirano12}. This superburst occurred at an accretion rate of less than 1\% of the Eddington rate, which is different from what has been observed from other sources. Thus, the EXO1745$-$248 event provides a valuable opportunity for sensitive studies of nucleosynthesis in X-ray superbursts. The superburst had an e-folding time of 6--11 hr, and the total emitted energy of 2--9 $\times 10^{42}$ erg assuming a distance of 5.5 kpc \citep{Ortolani07}. Using the superburst lightcurve modeling by \citet{cumming06}, the ignition column depth was estimated to be about $2 \times 10^{12}$ g cm$^{-2}$\citep{altamirano12,serino16}. The companion of EXO 1745$-$248 has not been identified, but based on the results of X-ray broadband spectral analysis, it is thought to be a ultracompact X-ray binary \citep{heinke03}. Due to the low accretion rate before the superburst onset, it is difficult to explain the origin of this particular event by carbon ignition, so the possibility of the He ignition origin has been pointed out \citep{altamirano12}. Moreover, the superburst was followed, 30 hr after its onset, by a luminous ($\sim$0.1 times the Eddington limit) outburst. The outburst may have been caused by an increase in the accretion rate, when the companion's photosphere, irradiated by the superburst, expands or evaporates. \citep{serino12}. In this paper, we report our finding of a strong X-ray emission structure at 6.6 keV in an X-ray spectrum of this object, obtained by RXTE at the rising phase of the outburst.

\begin{figure*}
 \begin{center}
  \includegraphics[width=17cm]{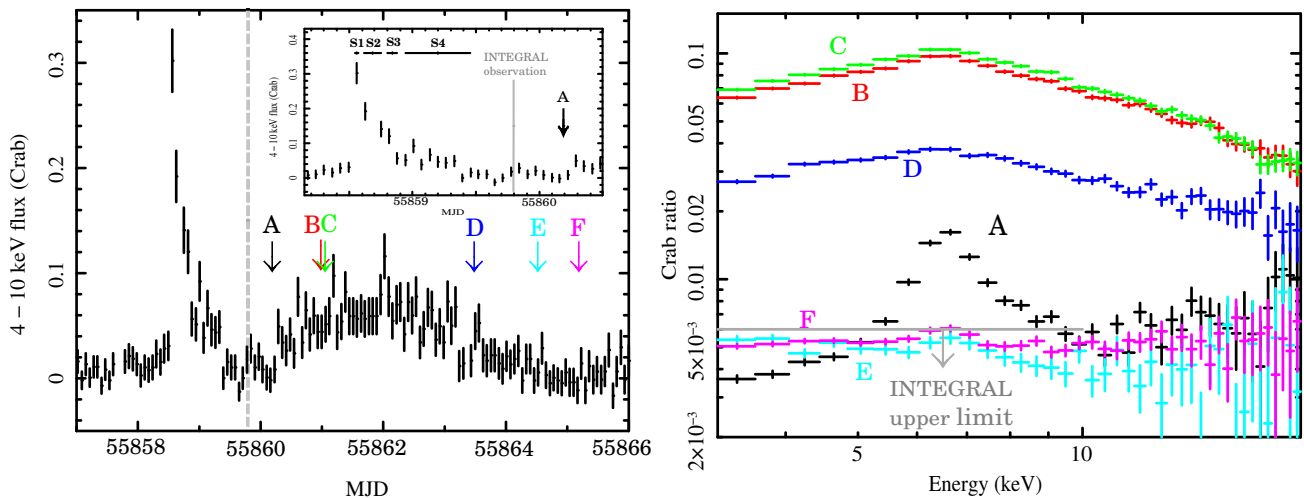}
 \end{center}
 \caption{(Left) MAXI/GSC lightcurve of EXO 1745$-$248 in the 4--10 keV band from MJD 55857 (2011 October 23) to 55868 (2011 November 3). The dashed line shows the INTEGRAL observation, whereas the arrows indicate the 6 RXTE observations. The inset shows a zoom of the lightcurve around the superburst. (Right) Energy spectra of EXO 1745$-$248 obtained by the RXTE/PCA, divided by that of the Crab nebula (see text). The black, red, green, blue, cyan and magenta specify the observations A to F defined in Table. \ref{table1}, respectively. The arrow indicates the INTEGRAL 5$\sigma$ upper limit of 6 mCrab in the 3--10 keV band \citep{vovk11}.}\label{fig1}
\end{figure*}

\section{Observations and Data reduction}
We used 6 data sets of EXO 1745$-$248, obtained with the RXTE/PCA \citep{jahoda06} from MJD 55860 (about 40 hr after the superburst onset) to 55865, and label them Obs.A to Obs.F. The log of these observations is given in Table \ref{table1}. The RXTE/PCA is composed of five Xe gas detectors (PCUs). The energy resolution of RXTE/PCA is 18\% at 6 keV, so it has little ability to resolve florescent X-ray lines with close centroid energies, but its large effective area (net geometric area per PCU is $\sim$1600 cm$^{2}$) allows us to obtain energy spectra with good statistics in a short exposure. The RXTE/PCA data were processed with the standard procedure using the CALDB files version 20120110. Lightcurves and spectra were extracted from the standard-2 data using the top layer of PCU-2. The background data of Obs.A, Obs.E and Obs.F were created using the FTOOL \texttt{pcabackest} with the faint source background model, whereas those of Obs.B, Obs.C, and Obs.D using \texttt{pcabackest} with the bright source background model.
Below, all errors are single parameter 1$\sigma$ errors.

\section{Data Analysis and Results}
Figure \ref{fig1} (left) shows the MAXI lightcurve of EXO 1745$-$248, where arrows indicate the timing of the 6 RXTE observations. As already reported by \citet{altamirano12}, the superburst started on MJD 55858.6 and lasted about 1 day. After the superburst ended, a rebrightening started at about MJD 55860 and lasted for about 5 days. This is the outburst that followed the superburst with an intermission interval of $\sim$30 hr \citep{serino12}. The RXTE observation timings can be classified into three epochs; in the rising phase (Obs.A), during (B, C, and D), and after (E, F) the outburst. 

Figure \ref{fig1} (right) shows the ratio of individual spectra to that of the Crab nebula (pulsar plus nebula), using a model based on the RXTE/PCA \citep{kuu03} as a template for the Crab spectrum. Surprisingly, in the Obs.A spectrum, a strong broad emission line structure is clearly seen around 6.6 keV. Hereafter, we refer to it as the Unusual Emission Structure, or ``UES''. In addition to these RXTE observations, the source was observed with INTEGRAL just 10 hr prior to Obs.A (and 29.7 hr after the MAXI superburst detection). However, no significant X-ray emission was detected, with an estimated 5$\sigma$ upper limit of 6 mCrab in the 3--10 keV band \citep{vovk11}. The MAXI data between the INTEGRAL observation and Obs.A indicate that there was no significant emission either, with an estimated 5$\sigma$ upper limit of 35 mCrab in the 4--10 keV band.

\subsection{Empirical modeling of the Obs.A spectrum}
\begin{figure*}[t]
 \begin{center}
 \includegraphics[width=17cm]{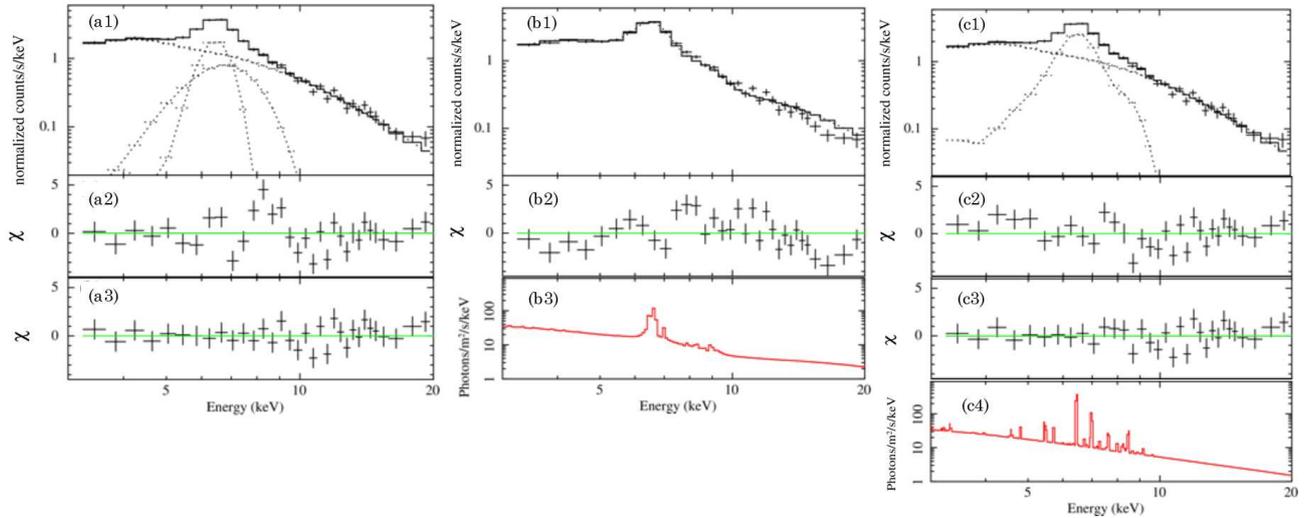}
\end{center}
\caption{(a1) The background-subtracted Obs.A spectrum of EXO 1745$-$248 obtained with the RXTE/PCA, fitted with an absorbed power law plus broad and narrow Gaussians. (a2) Residuals when only a broad Gaussian is used to express the UES. (a3) Residuals from the fit in panel (a1), using a power law and two Gaussians. (b1) The same RXTE spectrum as in panel (a1), fitted with an ionized reflection model \texttt{xilconv}. (b2) Residuals from the fit in panel (b1). (b3) The best-fit \texttt{xilconv} model determined in (b1). (c1) The same spectrum, fitted with an absorbed power law and redshifted CX emission. The abundances of Ti, Cr, and Co were allowed to vary, whereas those of the other elements were fixed to their solar values. (c2) Residauals when only the Cr abundance is allowed to vary. (c3) Residuals corresponding to (c1), which is the best-fit result . (c4) The best-fit CX model determined in (c1), shown in its incident form.}\label{gauss_cx_fit}
\end{figure*}

To quantify the UES, we tried to fit the Obs.A spectrum first using an absorbed power law for the continuum, and a broad Gaussian emission model for the UES. The hydrogen column density was fixed at $2.4 \times 10^{22}$ cm$^{-2}$ as determined from modeling hard-state spectra of this source obtained with the Swift/XRT \citep{parikh17}, and the Gaussian parameters (centroid energy, width, and normalization) were left free. However, as shown in figure \ref{gauss_cx_fit} (a2), this power law plus Gaussian model failed to reproduce the Obs.A spectrum ($\chi^2$/d.o.f = 87.42 / 28 with a null hypothesis probability of 5.0$\times 10^{-8}$) because of prominent residuals in the 5--9 keV region.
Therefore, next we added a narrow Gaussian emission component to better describe the UES, tying together the centroid energies of the two Gaussian components. As shown in figure \ref{gauss_cx_fit} (a1) and (a3), the UES is well reproduced ($\chi^2$/d.o.f = 32.5 / 27 with a null hypothesis probability of 0.2) by this power law plus double Gaussain model. If we untie the 2 centroid energies, the best-fit centroid energy of the narrow and broad Gaussians became 6.55$^{+0.02}_{-0.01}$ keV and 6.71$^{+0.07}_{-0.08}$ keV, respectively. Since the chance probability of this improvement by untying the two centroid energies is 0.16 as determined by \texttt{ftest} in XSPEC, hereafter we tie them. As summarized in table \ref{table2}, the best-fit model has a Gaussian centroid energy of $\sim$6.6 keV, which is close to the energy of He-like Fe K$\alpha$ line but lower by 0.1 keV. To evaluate the Equivalent Width (EW) of the UES, we used the formula generally defined as
\begin{equation}
\mathrm{EW} = \int^{\mathrm{10~keV}}_{\mathrm{4~keV}} \frac{F(E) - F_c(E)}{F_c(E)} ~dE 
\end{equation}
where $F(E)$ is the total flux density including the UES and $F_c(E)$ is the flux densiy in the continuum. The integration range of 4 - 10 keV covers the broad UES structure. We then obtain EW=4.3 $\pm$ 0.2 keV, which implies a line photon flux of $4.7 \times 10^{-3}$ ph cm$^{-2}$ s$^{-1}$.

We also estimated the upper limit of the UES in the INTEGRAL non-detection observations performed 10 hr before Obs.A. As is clear from figure \ref{fig1}, INTEGRAL would have certainly detected the UES if its flux were as strong as in Obs.A. Assuming the same continuum flux as in Obs.A, the line photon flux during the INTEGRAL observation must have been less than 1.5 $\times$ 10$^{-3}$ ph cm$^{-2}$ s$^{-1}$, or one third of that in Obs.A, in order to meet the INTEGRAL upper limit of 6 mCrab. If we alternatively assume the extreme case where the flux of the continuum is zero, then the upper limit of the UES line photon flux during the INTEGRAL observation is 9.7 $\times$ 10$^{-3}$ ph cm$^{-2}$ s$^{-1}$, or twice that in Obs.A. We also utilize the MAXI 5$\sigma$ upper limit of 35 mCrab in 4--10 keV, obtained between the INTEGRAL observation and Obs.A. It means that the UES flux during this period was less than 34.9 $\times$ 10$^{-3}$ ph cm$^{-2}$ s$^{-1}$, or eight times that in Obs.A.

Since the energy resolution of the RXTE/PCA does not allow us to determine whether the UES is an extended structure or a superposition of narrow emission lines, we investigated whether the Obs.A spectrum could be reproduced with a superposition of narrow Gaussians, instead of a broad Gaussian. When fitted with an absorbed power law plus three or four narrow Gaussian models, the chi-square values (null hypothesis probability) are $\chi^2$/d.o.f = 40.1 / 25 (0.028) and $\chi^2$/d.o.f = 29.2 / 23 (0.17), respectively. Therefore, the fit goodness of the four narrow Gaussian model is similar to that with the broad+narrow Gaussian modeling. In this case, the Gaussian centroid enegies were 5.5$^{+0.2}_{-0.3}$, 6.5$^{+0.1}_{-0.1}$, 7.5$^{+0.2}_{-0.2}$ and 8.6$^{+0.3}_{-0.2}$ keV, together with the Gaussian normalization of 0.5$^{+0.1}_{-0.2}$, 3.0$^{+0.1}_{-0.1}$, 0.7$^{+0.3}_{-0.1}$ and 0.3$^{+0.1}_{-0.1}$, respectively, all in units of $ 10^{-3}$ ph cm$^{-2}$ s$^{-1}$.

The 3--20 keV unabsorbed flux of the Obs.A power law component, estimated from the best-fit results, was $1.5 \times 10^{-10}$ ergs cm$^{-2}$ s$^{-1}$. Assuming a distance of 5.5 kpc \citep{Ortolani07}, the flux yields, in the same energy band, an X-ray luminosity 5.3$ \times 10^{35}$ ergs s$^{-1}$. Thus, the source was in the Low/Hard state during Obs.A. According to some studies of NS-LMXB \citep[e.g.,][]{sakurai2012,sakurai2014,wijnands15}, the observed continuum photon index of $1.8$ is typical in the Low/Hard state, but the UES is an unusual feature. Compared with the quiescent luminosity range of $3 \times 10^{31} - 2 \times 10^{34}$ erg s$^{-1}$ in the 0.5-10 keV band observed by Chandra \citep{rivera18}, the source was more luminous. Therefore, Obs.A was in the rising phase of the outburst.

\begin{table*}[t!]
  \tbl{Best-fit parameters for the Obs.A spectrum using a power law model.}   {%
 \begin{tabular}{lccc}
  \hline\noalign{\vskip3pt} 
  \hline\noalign{\vskip3pt} 
   & power law + two Gaussian   &power law + CX &power law + CX\\
  & & ($v_\mathrm{co}=10^{3}$km s$^{-1}$) & ($v_\mathrm{co}=10^{5}$km s$^{-1}$) \\
   \hline
$\Gamma$& 1.80 $^{+0.03}_{-0.03}$ & 1.82 $^{+0.04}_{-0.11}$ & 1.77$^{+0.04}_{-0.03}$ \\
$A^{*}$  &1.58$^{+0.03}_{-0.03}$ & 1.57 $^{+0.03}_{-0.03}$ & 1.52$^{+0.04}_{-0.05}$\\
$E^{\dagger}$ (keV)& 6.58 $^{+0.02}_{-0.02}$&  &\\
$W_{\mathrm{broad}}^{\ddagger}$ (keV)& 2.7 $^{+0.2}_{-0.2}$ & --- & ---\\
$A_{\mathrm{broad}}^{\S} $ &2.5$^{+0.2}_{-0.1}$ & ---& ---\\
$A_{\mathrm{narrow}}^{\S}$  &2.2 $^{+0.2}_{-0.3}$ & ---& ---\\
Redshift $z$ &--- & 0.037$^{+0.004}_{-0.004}$& 0.046$^{+0.005}_{-0.013}$\\
$kT$ (keV) $^{\|}$ & --- & 4.3$^{+1.3}_{-1.1}$& 2.9$^{+0.3}_{-0.2}$\\
$n_\mathrm{H}n_\mathrm{Fe}V$  (cm$^{-3}$) $ ^{\#}$ & ---& 1.3$^{+1.8}_{-0.4}$ & 1.2$^{+0.2}_{-0.3}$\\
Abundance ratio (Ti/Fe) & ---& 0.29$^{+0.08}_{-0.12}$ &0.75$^{+0.15}_{-0.16}$\\
Abundance ratio (Cr/Fe) & ---& 0.65$^{+0.12}_{-0.08}$& 2.3$^{+0.3}_{-0.5}$\\
Abundance ratio (Co/Fe) & ---& 1.8$^{+0.3}_{-0.5}$& 13.9$^{+1.2}_{-2.9}$\\
$\chi^2$ / d.o.f &32.5/27 & 31.8/25 & 32.2/25\\
\hline
\end{tabular}}
\label{table2}
\begin{tabnote}
The hydrogen column density was fixed at 2.4$\times$10$^{22}$ cm$^{-2}$. The width of the narrow  Gaussian emission component was fixed to zero. Both centroid energies of the Gaussian model are tied together. All errors denote 1$\sigma$ error level.\\
$*$ The 3 - 20 keV flux of the power law in units of 10$^{-10}$ erg cm$^{-2}$ s$^{-1}$. \\
$\dagger$ Centroid energy of the Gaussian.\\
$\ddagger$ FWHM of the line width of the broad Gaussian model.\\
$\S$ Normalization of the Gaussian. Defined in units of 10$^{-3}$ photons keV$^{-1}$ cm$^{-2}$ s$^{-1}$. \\
$\|$ The ionization balance temperature used in the calculation. The CX collision velocity is set to 1000 km s$^{-1}$.\\
$\#$ The emission measure $n_\mathrm{H}n_\mathrm{Fe}V$ in units of 10$^{54}$ (cm$^{-3}$), where the $n_\mathrm{H}$ is the neutral hydrogen density, $n_\mathrm{Fe}$ is the ionized iron density and $V$ is the effective interaction volume.
\end{tabnote}
\end{table*}

\subsection{Physical modeling of the Obs.A spectrum}
\subsubsection{Reflection model}
Although the UES is likely to be an atomic feature created by some 
Fe-group element(s), its width and EW are both unusually large, and need explanations. 
Since broad Fe K emission lines due to X-ray reflection off an accretion disk 
have sometimes been reported from LMXBs \citep{cackett2010,cackett12}, 
another possible explanation of the UES is 
that it is caused by reflection of the continuum X-rays off some thick materials.
Actually, \citet{matranga2017} detected a broad Fe line from EXO 1745$-$248. 
However, these broad Fe-K lines have typical EWs of only 100--300 eV,
which are much smaller than that of the UES, 4.3 keV. 
The EW of the reflected Fe-K line will increase when the direct component is completely blocked by thick materials, only the reflected component is visible, and Fe is nearly neutral \citep{Garcia2013}.
In this extreme case,
the EW would reach about 2 keV for solar Fe abundance
\citep{makishima1986, Garcia2013},
as determined by the ratio between the photoelectric and Thomson cross sections.
Therefore, the reflection model might explain the UES
if the reflector has an Fe abundance which is several times solar value.

Even if invoking a high Fe abundance,
another important difference remains:
the previously observed broad Fe-K lines  
extend mainly on the red side,
whereas the present UES exhibits both red and blue wings,
without evidence of an Fe-K absorption edge on the latter.
This difference might be solved
if assuming Fe to be highly ionized,
because the sharp Fe K-edge structure would be then suppressed.
To test this case, 
we fitted the Obs.A spectrum with an ionized reflection model \texttt{xilconv} \citep{Done2006,Kolehmainen2011} in XSPEC, 
which is a convolution model incorporating an ionized disk table from the XILLVER code \citep{Garcia2013}. 
We also assumed the aforementioned extreme geometry, together with a high Fe abundance.
Figure \ref{gauss_cx_fit} (b1) shows the derived fit.
Although the EW is approximately reproduced 
by invoking a Fe abundance higher than four times solar,
the model in fact fails ($\chi^2$/d.o.f of 90.2 / 29)
to reproduce the UES profile, mainly on the blue side (around 8 keV; panel b2).
In particular, a smeared edge feature at $\sim10$ keV,
presumably due to H-like Fe ions (see panel b3),
degrades the continuum fit above 15 keV.

Finally, the UES might still be explicable as a reflection,
if we considered strong relativistic effects, including light bending,
that may take place in an optically-thick accretion disk
extending down to a close vicinity of the NS.
However, in the Low/Hard state, the disk is known 
to be truncated at several tens kilometers from the NS \citep{ono2017}, 
and would not satisfy the required physical condition.
We therefore  conclude that the UES is not an Fe emission line due to reflection, 
and no longer consider this interpretation in the rest of the paper.

\subsubsection{Radiative recombination continuum model}
Since some heavy-element ashes that are highly ionized by the burst emission can be ejected in the form of radiation-driven winds during the PRE phase and fall back to the NS, we considered next Radiative Recombination Continuum (RRC) of ionized heavy elements as another possible explanation of the UES. An obvious advantage of this interpretation is that the expected feature can be further broadened by incorporating several different elements. Using an emission model from non-equilibrium ionization plasma \citep{kaastra1993}, we found that the UES can in fact be roughly represented with $\chi^2$/d.o.f of 49.3/30 by the sum of three redshifted RRC components from fully ionized Ca, Ti and Cr, with a redshift of $z\sim0.02$. However, the data do not show significant evidence of Lyman series cascade lines of these atoms; for example, H-like Ca k$\alpha$ emission line should appear at around 4 keV with an EW of $\sim$300 eV, whereas there is no significant emission structure with the upper limit of 100 eV. Then, the line-forming region would have to be under a special condition, where only the lines are strongly absorbed in contrast with the strong RRCs. Actually, the cascade lines can be self-absorbed by a slab of ionized material of the same species, while RRCs may penetrate farther than lines in an atmosphere with a certain range of densities. We studied this condition quantitatively, and found that the EW of the H-like Ca k$\alpha$ emission line becomes less than 100 eV only when the atmospheric column density is in a narrow range of 2 - 8$\times 10^{18}$ cm$^{-2}$. If the column density was higher than this, an absorption line with EW greater than 100 eV should be observed, in disagreement with the observed Obs.A spectrum. We consider that such a fine tuning is unnatural, and hence disfavor this interpretation. 

\subsubsection{Charge exchange reaction model}
Finally, we considered the charge exchange (CX) reaction between an ionized plasma (e.g., fall back burst winds) and a cool material (e.g., an accretion disk). We tried to fit the Obs.A spectrum using an absorbed power law plus the CX model provided by the SPEX code \citep{gu2016}, again with the intrinsic redshift factor $z$. Although we do not know the relative velocity $v_\mathrm{co}$ for collisions between the ionized plasma which accepts electrons and the neutral material supplying the electrons, we assume here $v_\mathrm{co} = $ 10$^3$ km s$^{-1}$, since the typical astrophysical velocity range is $\sim$100 - 5000 km s$^{-1}$ as discussed in \citet{gu2016}. The normalization of the CX process is directly proportional to the emission measure $n_\mathrm{H}n_\mathrm{Fe}V$, where $n_\mathrm{H}$ is the neutral hydrogen density, $n_\mathrm{Fe}$ is the number density of the ionized Fe ions, and $V$ is the effective interaction volume. In the fitting, we fixed the Fe abundance of the ionized plasma to the solar value, and consider the element abundances all relative to Fe. This is because we cannot estimate the plasma's ion density, which would be necessary to define the absolute abundances of the other elements.

First, we assumed that the relevant heavy elements in the CX plasma have the solar abundances relative to Fe. The fitting result shows that the He-like Fe CX lines with $z=$0.06 mainly contribute to the UES, but this case failed ($\chi^2$/d.o.f = 362.9/28) to reproduce the Obs.A spectrum with residuals around 5--9 keV. 
Since the model fitting results using the four narrow Gaussians indicate the presence of the emission line around 5.5 keV,
next we varied the Cr abundance (again relative to Fe) as a free parameter. Then, by adding CX lines from He-like Cr, to those from Fe, both with $z=0.02$, improved the fit significantly with $\chi^2$/d.o.f = 47.8/27. As shown in figure \ref{gauss_cx_fit} (c2), however, prominent residuals are still visible around 4--5 keV and 7--9 keV, so we further considered contributions of the CX lines from Ti, V, Mn, Co, Ni, Cu and Zn. Then, as shown in figure 2 (c3), the Obs.A spectrum has been reproduced successfully ($\chi^2$/d.o.f = 31.8/25), when z$=$0.037 is employed, and the Ti/Fe, Cr/Fe, and Co/Fe abundance ratios are allowed to take values of 0.29$^{+0.08}_{-0.12}$, 0.65$^{+0.12}_{-0.08}$, and 1.8$^{+0.3}_{-0.5}$, respectively, or 101$^{+26}_{-40}$, 42$^{+8}_{-5}$ and 640$^{+117}_{-170}$ times higher than the solar ratios. The inferred incident model is shown in figure \ref{gauss_cx_fit}(c4), and the best-fit parameters are summarized in table \ref{table2}. The upper limits on the V/Fe, Mn/Fe, Ni/Fe, Cu/Fe and Zn/Fe ratios, with the other parameters fixed to those in table \ref{table2}, are 0.21, 0.29, 0.57, 0.68 and 0.53, respectively. Assuming a NS mass of 1.4 $M_{\odot}$, the best-fit redshift of $z=$0.037 implies a distance of $\sim$60 km from the NS surface if z is a gravitational redshift. Thus, the CX modeling suggests an attractive possibility of strongly enhanced abundances of Ti, Cr and Co. Since the CX cross section is correlated with $v_\mathrm{co}$, we repeated the CX model fitting assuming a more extreme case of $v_\mathrm{co} = 10^5$ km s$^{-1}$, which is the free-fall velocity near the NS. As summarized in table 2, this case was found to requires several times higher abundances of the heavy elements to reproduce the UES than in the case of $v_\mathrm{co} = 10^3$ km s$^{-1}$. The reality of this scenario is examined in Discussion.

\subsection{Time variations during Obs.A}
\begin{figure}[t]
 \begin{center}
  \includegraphics[width=7cm]{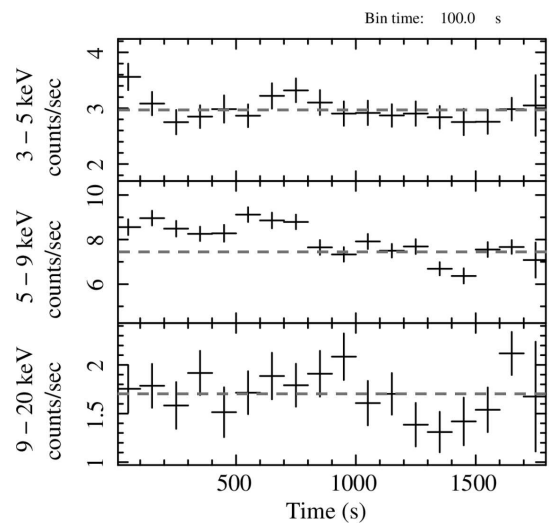}
 \end{center}
 \caption{Background-subtracted lightcurves of EXO 1745$-$248 obtained in Obs.A, in the 3--5 keV (top), 5--9 keV (middle) and 9--20 keV (bottom) bands.   }\label{fig3}
\end{figure}

To investigate time variations of the UES during Obs.A, we examined the background-subtracted lightcurves in the 3--5 keV (lower than the UES), 5--9 keV (the energy range including the UES), and 9--20 keV (higher than the UES) bands. The results given in figure \ref{fig3} suggest that the 5--9 keV intensity decreased through Obs.A, whereas those in the other two ranges are constant. In fact, when fitting a constant to the three lightcurves, the 3--5 keV, 5--9 keV and 9--20 keV data gave $\chi^2$ / d.o.f of 0.9, 5.6 and 1.0, respectively. Thus, the intensity of the UES varied independently of the continuum emission which was essentially constant during Obs.A. If we model the 5--9 keV lightcurve by a constant term plus an exponential decay function, the decay time constant was obtained as 1.0 hr. Here the constant term was fixed to the value predicted by the power law, that  gave the best fit to the 5--9  keV spectrum when combined with the two Gaussians.

\begin{table}
  \tbl{Line EWs and fluxes in the S1, S2, S3 and S4 spectrum during the superburst.}   {%
 \begin{tabular}{lcccccc}
  \hline\noalign{\vskip3pt} 
  \hline\noalign{\vskip3pt} 
   Obs. & S1 & S2  & S3 & S4 \\
   \hline
   EW (keV)$^{*}$ & $<$0.37 & $<$0.87 & $<$0.61 & $<$2.39  \\
   $A^{\dagger}$ & $<$38 & $<$54 & $<$20 & $<$16 \\
\hline
\end{tabular}}
\label{table_burstupper}
\begin{tabnote}
$*$ EW of the UES defined as equation (1).\\ 
$\dagger$ Normalization of the Gaussian. Defined in units of 10$^{-3}$ photons keV$^{-1}$ cm$^{-2}$ s$^{-1}$. \\
\end{tabnote}
\end{table}

\subsection{Searching for the UES in the other observation phases}
\subsubsection{During the superburst}
As shown in \citet{serino12}, the X-ray spectrum in the superburst obtained by MAXI was well reproduced by a blackbody model, and no UES-like emission line structure was detected. In order not to miss any transient emergence of the UES, we divided the spectral data of the MAXI superburst into four time intervals (S1-S4) shown in the inset of figure 1. The MAXI spectral data were obtained with the same data reduction as in \citet{serino12}, and the upper limits were obtained using a blackbody model for the continuum and the narrow and broad Gaussian emission model for the UES in Section 3.1. The derived results, shown in table \ref{table_burstupper}, do not strongly constrain the photon flux, but excludes the presence of emission line structures with the EW similar to that of the UES.

\begin{table*}
  \tbl{Line EWs and fluxes in the Obs.B, C, D, E and F spectrum.}   {%
 \begin{tabular}{lcccccc}
  \hline\noalign{\vskip3pt} 
  \hline\noalign{\vskip3pt} 
   Obs. & A & B  & C & D & E & F\\
   \hline
   EW$^*$ (keV) & 4.3$^{+0.2}_{-0.2}$ & 0.33$^{+0.02}_{-0.02}$ & 0.28$^{+0.02}_{-0.02}$ & 0.15$^{+0.04}_{-0.04}$ & 0.25$^{+0.09}_{-0.09}$ & 0.19$^{+0.05}_{-0.05}$ \\
   $A^{\dagger}$ & 4.7$^{+0.2}_{-0.2}$ &6.0$^{+0.4}_{-0.4}$ &4.9$^{+0.5}_{-0.5}$ & 1.2$^{+0.3}_{-0.3}$ & 0.3$^{+0.1}_{-0.1}$ &  0.24$^{+0.06}_{-0.06}$\\
\hline
\end{tabular}}
\label{table4}
\begin{tabnote}
$*$ EW of the UES defined as equation (1).\\ 
$\dagger$ Normalization of the Gaussian. Defined in units of 10$^{-3}$ photons keV$^{-1}$ cm$^{-2}$ s$^{-1}$. \\
\end{tabnote}
\end{table*}

\begin{figure}[t]
 \begin{center}
  \includegraphics[width=7cm]{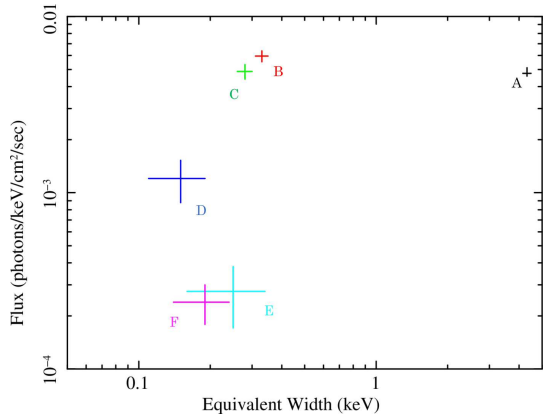}
 \end{center}
 \caption{A scattered plot between the best-fit values of the EW and photon flux from Obs.A--Obs.F referring to table 4. Both quantities are summed over the two Gaussians (for details see text).}\label{fig4}
\end{figure}

\subsubsection{During the outburst}
After Obs.A, the rebrightening due to the outburst started and lasted for 5 days (figure \ref{fig1}). Although the spectra from Obs.B--Obs.F are much more featureless than that of Obs.A (figure \ref{fig1} right), they all exhibit a weak emission structure around 6.5 keV. To examine whether these features in Obs.B--Obs.F are significantly smaller in EW than the UES in Obs.A, we fitted the Obs.B--Obs.F spectra in the 3--20 keV energy range, using the narrow and broad Gaussian emission model. The centroid energy, width, and the intensity ratio between the two Gaussians were fixed at the values from the Obs.A spectrum. To express the continuum in  Obs.B, C and D, we adopted an absorbed cutoff power law model with absorption edge, which mimics reflection of Comptnized hard X-rays from a disk. On the other hand, as for the fainter continuum in Obs.E and Obs.F, we adopted an absorbed power law model. The results are listed in table \ref{table4} and the derived EWs and the photon fluxes (both summed over the two Gaussians) are presented in a scatter plot in figure \ref{fig4}. The EWs in Obs.B--Obs.F are significantly smaller than for the UES in Obs.A, whereas the photon flux during Obs.B and Obs.C are almost the same as that of the UES. Therefore, from the spectrum alone, we cannot exclude the possibility that the UES was still present in Obs.B and Obs.C. However, considering the rather short decay time constant (1 hr; Section 3.3), it is reasonable to suppose that the UES had already decayed by Obs.B (17 hr after Obs.A). We conclude that the weak emission line around 6.5 keV in Obs.B--Obs.F is an ionized Fe line usually seen in EXO 1745$-$248, whereas the UES in Obs.A is distinct in origin. The weak Fe lines observed in Obs.B--Obs.F are essentially the same as the ionized Fe line previously observed at this energy with the EW of $\sim$0.1--0.2 keV, from EXO 1745$-$248 \citep{heinke03,matranga2017} and other LMXBs \citep{cackett12}. 
We cannot verify whether the weak emission line is present in Obs.A or not, but the EW of the line in Obs.B--F is about 5\% compared to that of the UES in Obs.A, thus it does not significantly affect the results of the spectral analysis of Obs.A in previous Section.

\section{Discussion}
\subsection{Summary of the observations}
We have presented the results on the UES observed with RXTE about 40 hours after the onset of the superburst. Superbursts at low accretion rates are very rare phenomena (only two superbursts from two sources in 11 years of MAXI observations; \citet{serino16,intzand17}), and such UES features have never been observed from this object (or other similar objects), even though many observations have been made by RXTE, Chandra, XMM-Newton, and INTEGRAL \citep{heinke03, wijnands2005, degenaar2012, matranga2017,rivera18}. Considering that the probability of these rare events occurring by chance within 40 hours of each other is very low, it would be natural to assume that the UES was associated with the superburst. In addition, not only the most successful CX modeling, but also the less acceptable RRC and reflection scenarios, all require considerable enrichment of the heavy elements to explain the UES, and this result also suggests a connection between the UES and the preceding superburst. Moreover, the fact that the UES was observed during the rising phase of the outburst suggests that the UES is associated with an increase in the accretion rate that may be an aftermath of the superburst.

Before trying to interpret the UES, let us briefly summarize its observed properties.
\begin{enumerate}
    \item  The UES was detected 39.6 hr after the superburst onset. At that time, the source was in a start phase of the outburst, and showed a 3--20 keV luminosity of 5.3$ \times 10^{35}$ ergs s$^{-1}$. However, the UES was not seen in the $<$ 6 mCrab INTEGRAL X-ray spectrum obtained 29.7 hr after the superburst onset. 
    \item The UES is centered at 6.6 keV, is broadened beyond the instrumental energy resolution, and is extremely strong with an EW of 4.3 keV.
    \item The UES was phenomenologically reproduced by a narrow + broad (2.7 keV in FWHM) Gaussian emission components both centered at 6.6 keV, or by four narrow Gaussian emission lines at 5.5, 6.5, 7.5 and 8.6 keV. 
    \item The UES was also reproduced by a physical model of CX emissions from Ti, Cr, Fe and Co, on condition that their abundance ratios to Fe are by about 1.5 to 3 orders of magnitude higher than the solar ratios.
    \item During the half an hour exposure time of Obs.A, the UES intensity decreased significantly, with an exponential time scale of 1 hr, whereas the continuum intensity was constant. The UES is likely to have disappeared by Obs.B (60 hr after the superburst onset).
\end{enumerate}
With this summary in mind, below we discuss possible origins of the emission feature. 
Any interpretation must be able to explain the four essential properties of the emission feature: (i) huge EW (4.3 keV), (ii) the time delay, (iii) the extreme line broadening, and (iv) the decay on a time scale of 1 hr.

\subsection{Light echo from circumstellar medium}
The simplest interpretation of the UES would be to regard it as a fluorescence line from iron, located somewhere in a vicinity of the system. Considering (ii), we examine a possibility that the UES is a light echo of the superburst emission from a dense circumstellar medium of the NS. 
Since the decay time scale of 1 hr is comparable to the superburst duration, the requirement (iv) is fulfilled. Indeed, this would provide an almost sole way to get around the required huge metal abundance. Then, to explain the requirement (i), we only need to assume that the circumsteller medium is uniformly distributed except the line of sight direction, with an absorption column density of $N_{\mathrm{H}}\sim5 \times$10$^{22}$ cm$^{-2}$ according to \citet{makishima1986}. As for the requirement (ii), the circumstellar medium must be located at a distance of $\sim$ 10$^{15}$ cm from the NS as determined by the 40 hr delay. If we accept such special situations, (i) and (ii) can be explained. However, we cannot explain (iii) because the emission line from the cold dense circumstellar medium must be narrow. Therefore, a light echo is unlikely as the origin of the UES.

\subsection{Electron capture by unstable heavy nuclei on the NS surface}
Since the UES is difficult to explain as Fe-K line emission only, we turn to possibilities that it involved other elements that were highly enriched by the superburst.  
One of such possibilities is that it arose via the electron capture process by unstable nuclei that were produced by the superburst. The burst-produced unstable nuclei are generally proton enriched, and will decay mostly either via $\beta^{+}$ decay process, or electron capture mainly from the K-shell. The latter process is followed by the emission of a K-line photon from a daughter nucleus, assuming that it has at least one L-shell electron. In this context, the observed centroid energies of the emission lines are expected to be redshifted by the gravitational field of the NS. Assuming a typical mass and radius of NSs (1.4 $M_{\odot}$ and 12 km, respectively), the observed 6.6 keV energy corresponds to a rest-frame energy of 8.0 keV. It is consistent with the K$\alpha$ line energy of Cu, which would accompany the electron capture decay of unstable Zn. 
However, this scenario cannot explain the non-detection of the signal in the INTEGRAL observation which started about 30 hr after the superburst peak time, because the line intensity should be decreasing monotonically from the superburst. Therefore, the electron capture scenario cannot provide  a plausible explanation, either. 

\subsection{Charge exchange emission between fall back burst winds and the accretion disk}

As found in Section 3, the UES in Obs.A was successfully fitted by a redshifted CX model. Here, we attempt to construct an astrophysical scenario which can explain the UES in terms of a CX reaction. 

The best-fit redshift of z=0.037$^{+0.004}_{-0.004}$ using a redshifted CX model imply that the emission region is located at a distance of $59^{+6}_{-5}$ km from the NS. Moreover, the abundance ratios of Ti/Fe, Cr/Fe and Co/Fe are 101, 42 and 640 times higher than the solar ratios, respectively. Therefore, the result suggests that some ionized plasma with very high heavy-element abundance collided, after the burst, with some neutral material at a distance of $\sim$60 km from the NS after the superburst. Then, what are the origins of the metal-enriched plasma and the neutral material?  Theoretical calculations by \citet{Yu2018} show that heavy-element ashes, such as $^{44}$Ti, $^{48}$Cr and $^{52}$Fe, made by thermonuclear X-ray bursts, can be ejected at their PRE phase in the form of radiation-driven winds. Evidence of such metal-enriched ejecta has been provided by several spectral absorption features detected in superexpansion bursts \citep{intzand10, kajava17}. Furthermore, \citet{barriere2015} reported a marginal detection of the absorption line of H-like Cr from a burst of GRS 1741.9$-$2853. Based on these studies, we suggests that the superburst launched a highly metal-enriched plasma.

Although the MAXI data provide no direct evidence for a PRE episode
in this superburst from EXO 1745-248, it has been reported that the peak of the superburst observed by Swift/BAT in the 15--50 keV band is significantly delayed relative to the MAXI peak in the 2--20 keV band \citep{altamirano12}. This delay provides evidence for a PRE phase,
because an expanded photosphere will reduced 
the burst temperature and hence suppress higher-energy X-rays,
until the photosphere comes back and  the temperature
returns to a value close to the local Eddington temperature \citep[e.g.,][]{tawara1984}.
Moreover, the low accretion rates before the superburst points to the possibility that this superburst was due to the He ignition \citep{altamirano12}. Then, the event is likely to have reached the Eddington limit, because bursts ignited in the He-rich layer are theoretically predicted to have high peak luminosities \citep[e.g.,][]{hanawa1982,woosley2004}. Therefore, it is reasonable to assume that this superburst reached the Eddington luminosity and generated a burst wind.

At the same time as the suggested PRE, the atmosphere of the 
companion star, evaporated by the superburst, began to accrete, causing the inner edge of the disk to move inward, leading to the outburst \citet{serino12}. Then, the accretion disk and the burst wind will interact each other during that transitional period, causing the CX emission.

To quantitatively examine the above CX hypothesis, first we investigated how much heavy elements mass can be ejected by a single superburst.
According to \citet{Yu2018}, who performed hydrodynamic simulations of the super-Eddington burst winds assuming the pure He accretion, the wind mass-loss rate $\dot{M}_\mathrm{w}$ and the wind duration $t_\mathrm{w}$ are estimated as
\begin{equation}
    \dot{M}_\mathrm{w} \simeq 2.1 \times 10^{18} y_{9}^{0.22} ~\mathrm{g~s^{-1}},
\end{equation}\
\begin{equation}
    t_\mathrm{w} \simeq 15 y_{9}^{0.79} ~\mathrm{s},
\end{equation}
where $y_{9}$ is the ignition column depth in units of 10$^9$ g~cm$^{-2}$. 
The present burst is thought to have had $y_{9} = 2 \times 10^3$, as shown in Section 1. Substituting it into equation (13) and (14), we obtain $M_\mathrm{w} = \dot{M}_\mathrm{w} \times t_\mathrm{w}$ $\simeq 7 \times 10^{22}$ g. \citet{Yu2018} also estimated that the mass fraction of the ejected heavy elements, such as $^{48}$Cr and $^{52}$Fe, is higher than 0.1 although they assumed $y_{9}$ = 0.3$-$5. Simply adopting the fraction of 0.1, the total mass of ejected $^{52}$Fe is estimated as $M_{\mathrm{Fe}} = 7 \times 10^{21}$ g. Because the half-life of $^{52}$Fe is 8.275 hr\footnote{This  half-lie applies to neutral $^{52}$Fe atoms, and provides a lower limit in the present context, because the life gets longer when the nucleus is ionized\citep[e.g.,][]{irnich95}. Thus, we assume the shortest life-time of  $^{52}$Fe here.}, the mass will be reduced to
\begin{equation}
M_{\mathrm{Fe}} \simeq 3 \times 10^{20}~\mathrm{g}
\end{equation}
in the 40 hr interval.

Next, we examine whether this $M_{\mathrm{Fe}}$ can be reconciled with the emission measure value,
\begin{equation}
n_\mathrm{H}n_\mathrm{Fe}V = 1.3\times 10^{54}    ~\mathrm{cm}^{-3},
\end{equation}
obtained using the physical CX model assuming $v_\mathrm{co} = $ 10$^3$ km s$^{-1}$ (Section 3). For this purpose, we assume;
\begin{itemize}
    \item The CX emission came from the accretion disk at a distance of $\sim$60 km from the NS, as indicated by the best-fit redshift value in table \ref{table2}.
    \item The requirement (iv) represents the decay time scale of $n_\mathrm{Fe}$ due to the CX recombination process.\footnote{One possibility is that the fast decay of the UES was caused by the decay of unstable nuclide $^{52}$Fe in the context of our CX emission hypothesis. However, the observed decay time constant of $\sim$1 hr is much shorter than the half-life of $^{52}$Fe (8.275 hr for a neutral $^{52}$Fe). Therefore, we suggest that the decay property is related to the recombination timescale for the ionization equilibrium although the quantitative discussion is beyond the scope of this paper. }
    \item The time profile of the UES is described by a fast rise and a simple exponential decay with a time constant of $\tau_\mathrm{d}$ = 1 hr.
    \item Throughout the CX emission, $n_\mathrm{H}$ and $V$ in equation (5) are constant.
    \item Over $V$, $n_{\rm Fe}$ is uniform.
\end{itemize}
Under these simple assumptions, $M_{\mathrm{Fe}}$ involved in the CX emission can be estimated as
\begin{equation}
M_{\mathrm{Fe}} = \int n_\mathrm{Fe}(t) ~V ~dt = \int n_\mathrm{Fe0}~e^{-\frac{t}{\tau_\mathrm{d}}}~V ~dt = n_\mathrm{Fe0} \times \tau_\mathrm{d} \times V.
\end{equation}
Here $n_\mathrm{Fe0}$ is the ionized iron density at the peak of the CX emission, which can be written as 
\begin{equation}
n_\mathrm{Fe0} = \frac{\mathcal{E}_0}{n_\mathrm{H}V},
\end{equation}
where $\mathcal{E}_0$ is the emission measure at that time. Since the upper limit on $\mathcal{E}_0$ obtained by MAXI between the INTEGRAL observation and Obs.A was eight times of equation (5), whereas it must be larger than equation (5), we obtain a constraint as
\begin{equation}
1.3\times 10^{54} < \mathcal{E}_0 < 1.0 \times 10^{55} ~\mathrm{cm}^{-3}.
\end{equation}
Let us try to convert this constraint, first into that of $n_\mathrm{H}$,
and then into an estimate of $M_{\mathrm{Fe}}$ via equation (6) and (7). 
According to \citet{kato08}, an accretion disk, assuming it to be dominated by gas pressure and electron scattering opacity, has a density as
\begin{equation}
\rho = 8.0 \times \alpha^{-7/10}m^{-7/10}\dot{m}^{2/5}\hat{r}^{33/20}f^{2/5} \mathrm{g/cm^{-3}},\\
\end{equation}
with
\begin{displaymath}
~m=M/M_{\odot},~\dot{m}=\dot{M}/(L_\mathrm{E}/c^2),~\hat{r}=r/r_{\mathrm{g}},~
f=1 - \sqrt{3r_{\mathrm{g}}/r},\\
\end{displaymath}
where $\alpha$ is the Shakura-Sunyaev viscosity parameter (which is hereafter assumed to be 0.1), $M$ is the mass of the NS, $\dot{M}$ is the accretion rate, $L_\mathrm{E}$ is the Eddington luminosity  ($L_\mathrm{E}/c^2$ is the critical mass-flow rate), $r$ is the radius along the disk, and $r_{\mathrm{g}}$ = $2GM/c^2$ is the Schwarzschild radius. At Obs.A, we can estimate as $\dot{M}~=$  1.2 $\times 10^{16}$ g s$^{-1}$ by extrapolating the observed 3--20 keV luminosity to the 0.1--200 keV band. At $r=60$ km, where the CX photons are thought to be emitted, we expect $\rho = 9.8 \times 10^{-2}$ from equation (9). Assuming the disk is composed of hydrogen, we obtain 
\begin{equation}
n_\mathrm{H} = 5.9 \times 10^{22} ~\mathrm{cm}^{-3}.
\end{equation}
Substituting equation (8) and (10) into equation (7), we obtain a constraint as
\begin{equation}
2.2\times 10^{31} < \mathrm{V} \times n_\mathrm{Fe0} \dot < 1.7 \times 10^{32} .
\end{equation}
Adopting the maximum condition of this equation, the total number of ionized Fe ions is estimated as 6.1 $\times~10^{35}$ from equation (6), which corresponds to a total Fe mass of 
\begin{equation}
M_{\mathrm{Fe}} = 5.7 \times 10^{13} ~\mathrm{g}. 
\end{equation}
Thus, even assuming the largest estimate for the required  ionized Fe mass, it is still some 7 orders of magnitude lower than the iron yield in the superburst, equation (4). In other words, the UES can be explained by the CX scenario as long as $\sim2 \times 10^{-7}$ of the total ejected mass falls back to the NS. 
According to the abundance ratio of Ti/Fe, Cr/Fe and Co/Fe by our spectral fitting using the CX model assuming $v_\mathrm{co} = $ 10$^3$ km s$^{-1}$ (Section 3 and Table 2), the total mass of ionized Ti, Cr and Co required to explain the UES are 1.7 $\times~10^{13}$ g, 3.7 $\times~10^{13}$ g and 1.0 $\times~10^{14}$ g, respectively. 

Finally, we consider the requirement (ii), i.e., the 40 hr delay. The free fall time from the distance of about 50 km is $\sim1\times10^{-3}$ s, which is far shorter than the requirement (ii). Although it is difficult to determine the cause of this 40 hr delay due to the limited information available, a possible scenario might be constructed if we focus on the very low accretion rate of EXO 1745-248, and the fact that the UES was observed during the rising phase of the outburst.
That is, the NS in this system is presumably rotating rapidly, with a certain level of magnetic field. When the accretion rate decreases after the superburst, the Alfv\`{e}n radius $R_\mathrm{A}$, where the gas pressure is equal to the magnetic pressure of the NS magnetosphere, will exceed the corotation radius $R_\mathrm{co}$, where the Keplerian frequency of the disk equals the stellar rotation frequency. Then, the matter will be prevented from accreting, by the propeller effect \citep[e.g.,][]{matsuoka2013}. The effect will continue until the revived accretion flow pushes the magnetosphere to make $R_\mathrm{A} < R_\mathrm{co}$.
To estimate whether this scenario works, we first evaluate $R_\mathrm{co}$, which is expressed as
\begin{eqnarray}
&R_\mathrm{co} = [GM_\mathrm{NS}P^2/(4\pi^2)]\nonumber\\
&=1.7\times10^{6}(M_\mathrm{NS}/1.4M_{\odot})^{1/3}(P/1~\mathrm{ms})^{2/3}~\mathrm{cm},
\end{eqnarray}\
where $M_\mathrm{NS}$ and $P$ are the mass and spin period of the NS, respectively. Although $P$ of EXO 1745$-$248 unknown, assuming a typical NS-LMXB period of 1-10 ms will give a constraint as
\begin{equation}
1.7 \times 10^{6} < R_\mathrm{co} < 7.9 \times 10^{6}~\mathrm{cm}. 
\end{equation}
Turning to $R_\mathrm{A}$, the formulation by \citet{GL79} and \citet{matsuoka2013} gives,
\begin{eqnarray}
    R_\mathrm{A} &= 4.8 \times 10^{6} \eta \left(\frac{L_X}{10^{36}~\mathrm{erg~s}^{-1}}\right)^{-2/7} \left(\frac{B}{10^{8}~\mathrm{G}}\right)^{4/7} \nonumber\\
    &\left(\frac{M_\mathrm{NS}}{1.4M_{\odot}}\right)^{1/7} \left(\frac{R_\mathrm{NS}}{1.2 \times 10^{6}~ \mathrm{cm}}\right)^{10/7}~\mathrm{cm},
\end{eqnarray}\
where $\eta=0.52-1, L$, and $B$ are the dimensionless parameter depending on the model of accretion flow, the luminosity, and the magnetic field strength, respectively.
Assuming $B = 10^8$ G, and employing the quiescent luminosity of EXO 1745$-$248, $3 \times 10^{31} - 2 \times 10^{34}$ erg s$^{-1}$ \citep{rivera18}, we obtain
\begin{equation}
7.6 \times 10^{6} < R_\mathrm{A} < 9.4 \times 10^{7} \mathrm{cm}.
\end{equation}
Comparing equation (14) and (16), an accretion rate as low as the quiescent state would make $R_\mathrm{A}$ larger than $R_\mathrm{co}$, and prevent any material, either the accretion flow or the fall-back burst wind, entering inside $R_\mathrm{A}$. The ionized burst wind will stay around $R_\mathrm{A}$.
When the mass accretion from the companion star increases \citep[e.g., due to the evaporation of the companion surface by the superburst,][]{serino12} and reaches $R_\mathrm{A}$ in 40 hr after the peak of the superburst, the propeller effect becomes ineffective due to the reversed condition as $R_\mathrm{A} < R_\mathrm{co}$. As a result, the burst wind and the accretion disk start falling to the NS, and coexist at a distance of about 60 km from the NS, which may have produced the CX emission. 
At this stage, the restarted mass accretion is considered to have had a rate of 
\begin{equation}
\dot{m} < 3\times10^{15}~\mathrm{g~s^{-1}},
\end{equation}
as constrained by the INTEGRAL observation between the superburst and Obs.A.
Therefore, it is likely that material was not accreted at a higher accretion rate than equation (17) from the end of the superburst until the period between Obs.A. Comparing this result with equation (4), it would take more than 28 hours for all the iron ejecta to accrete to the NS. Since this result is comparable to the observed 40 hr delay, the scenario may also meet the requirement (iv). To further test this CX scenario, it will be important to observe UES with detectors that have better energy resolution than gas detectors, for example NICER \citep{gendreau2016} or XRISM \citep{tashiro2018}.

\section{Conclusion}
In an X-ray spectrum of EXO 1745$-$248 acquired with RXTE,
we discovered an Unusual Emission Structure (UES), 
which is centered at 6.6 keV, is broadened, and has an unusually large EW of 4.3 keV.
It was detected in 2011 October, 40 hr after a thermonuclear superburst,
when the source was about to enter into the subsequent accretion outburst.
Throughout the 1.7 ks observation, the UES decreased 
significantly  in intensity with an e-folding time of 1 hr, 
while it was undetectable either 10 hr before or 20 hr after this observation.
Empirically, the UES was reproduced by a narrow + broad (2.7 keV in FWHM) Gaussian emission components both centered at 6.6 keV,
or by four narrow Gaussian emission lines at 5.5, 6.5, 7.5, and 8.6 keV.
We also tried to fit the UES spectrum with  several physical models.
Although the broad-Fe line models failed to give a reasonable account,
the CX model invoking K-shell emission from Ti, Cr, Fe, and Co was successful, 
as long as these metals are highly enriched and are redshifted by $z=0.037$.
To explain these observational results, we constructed a scenario,
which may provide some clues to the nucleosynthesis in these objects.
(1) A certain amount  of highly ionized Fe ions were synthesized
 in the superburst, and ejected from the NS as burst winds.
(2) The reduced accretion rate after the superburst made the Alfven radius
of the NS exceed  its co-rotation radius, so that the propeller effect 
of the NS evacuated the region inside the Alfven radius.
(3) After a while, the mass accretion started, compressed the Alfven shell,
and suppressed the propeller effect,
to allow the wind plasma to fall back to the NS.
(4) The plasma encountered the accretion disk moving inward
at $\sim$60 km from the NS, where the CX photons were produced.

\bigskip 
\begin{ack}
The authors appreciate very much the many constructive comments from the anonymous referee. We also gratefully acknowledge discussions with Takayuki Yamamoto. This research was supported by the Special Postdoctoral Researchers Program in RIKEN (WI), Japan Society for the Promotion of Science (JSPS) KAKENHI Grant Numbers JP16K17717 (WI), JP20H04743 (WI), JP19H00704 (HY), JP20H00175 (HY), JP19K14762 (MS).
\end{ack}


\bibliographystyle{aasjournal}
\bibliography{ref.bib}

\end{document}